\documentclass[aps,prd,showpacs,amssymb,superscriptaddress, notitlepage,floats,floatfix,nofootinbib, twocolumn]{revtex4-1}

\usepackage{graphicx}
\usepackage{dcolumn}
\usepackage{bm}
\usepackage{amsmath}
\usepackage{color}
\usepackage{hyperref}
\usepackage{journals}
\usepackage{wasysym}
\usepackage[normalem]{ulem}

\definecolor{darkgreen}{rgb}{0.0, 0.6, 0.0}

\begin{document}

\title{Bayesian Inverse Problem of Rotating Neutron Stars}

\author{Sebastian H. V\"olkel}
\email{sebastian.voelkel@sissa.it}

\affiliation{SISSA, Via Bonomea 265, 34136 Trieste, Italy and INFN Sezione di Trieste}
\affiliation{IFPU - Institute for Fundamental Physics of the Universe, Via Beirut 2, 34014 Trieste, Italy}

\author{Christian J. Kr\"uger}
\email{christian.krueger@tat.uni-tuebingen.de}

\author{Kostas D. Kokkotas}
\email{kostas.kokkotas@uni-tuebingen.de}

\affiliation{Theoretical Astrophysics, IAAT, University of T\"ubingen, 72076 T\"ubingen, Germany}

\date{\today}

\begin{abstract}
In this work we provide a framework that connects the co-rotating and counter
rotating $f$-mode frequencies of rotating neutron stars with their stellar
structure. The accurate computation of these modes for realistic equations of state has
been presented recently and they are here used as input for a Bayesian
analysis of the inverse problem. This allows to quantitatively reconstruct
basic neutron star parameters, such as the mass, radius, rotation rate or
universal scaling parameters. We find that future observations of both
$f$-mode frequencies, in combination with a Bayesian analysis, would provide a
promising direction to solve the inverse stellar problem. We provide two
complementary approaches, one that is equation of state dependent and one that
only uses universal scaling relations. We discuss advantages and disadvantages
of each approach, such as possible bias and robustness. The focus is on
astrophysically motivated scenarios in which informed prior information on the
neutron star mass or rotation rate can be provided and study how they impact
the results.
\end{abstract}

\maketitle

\section{Introduction}\label{introduction}

Ongoing advances in gravitational wave astronomy offer unprecedented
opportunities to study the complex and rich physics of neutron stars. The
milestone detection GW170817 of two merging neutron stars
\cite{2017PhRvL.119p1101A,2017ApJ...848L..12A,2017ApJ...848L..13A} (and
subsequent events such as GW190425 \cite{2020ApJ...892L...3A}) triggered
an enormous amount of studies and insights to a wide range of questions on
neutron stars, nuclear physics, cosmology, and fundamental physics~\cite{2017ApJ...850L..19M,2017ApJ...850L..34B,2017PhRvL.119y1301B,2017PhRvL.119y1303S,2017PhRvL.119y1304E,2018ApJ...852L..29R,2018PhRvL.121p1101A}. Among the
key observables that should be detectable in the future by advanced detectors
are the $f$-mode frequencies of massive post-merger objects or rotating
neutron stars (both isolated as well as part of a binary system). Those are a
promising laboratory to study extreme nuclear physics and strong field gravity
in more detail and will help constrain the nuclear equation of state
(henceforth EOS)~\cite{2014PhRvD..90b3002B, 2014PhRvL.113i1104T}.

While binary mergers involving at least one neutron star will provide a
plethora of observations that can be utilized in order to put constraints on
current unknowns in nuclear physics, we will in this study focus on the
sub-field of \emph{gravitational wave
asteroseismology}~\cite{1996PhRvL..77.4134A, 1998MNRAS.299.1059A,
2005ApJ...629..979L}. Starting in the 1970s, helioseismology has proven
extremely successful in gaining a highly detailed knowledge on the composition
as well as processes operating in the interior of the Sun, based on the
observation of acoustic modes visible on the solar
surface~\cite{1976Natur.259...89C, 2002RvMP...74.1073C}. Asteroseismology of
neutron stars will never be able to reproduce that level of detail known from
solar studies, however, it will play a crucial role in the long-lasting
effort to constrain the nuclear EOS; the inverse problem has to be tackled.
Even though neutron stars exhibit a rich spectrum of different oscillation
modes~\cite{1941MNRAS.101..367C, 1985ApJ...297L..37M, 1992MNRAS.255..119K}, we
will focus on the fundamental acoustic mode, or more precisely, the $l=|m|=2$
$f$-mode, as it is the fluid mode strongest coupled to gravitational radiation
and hence most likely to be detected via future gravitational wave
observations. The computation of the $f$-modes of arbitrarily fast rotating
neutron stars is numerically very complex and has a long history
\cite{1967ApJ...149..591T, 1983ApJS...53...73L, 2002MNRAS.334..933J,
2008PhRvD..78f4063G, PhysRevD.102.064026}. Asteroseismology often relies on
the availability of certain \emph{universal relations} that are independent of
the nuclear EOS, i.e., they allow deductions from mode frequencies to bulk
properties of the observed star by solving the inverse problem. Several such
universal relations have been proposed for various oscillation modes of
neutron stars; they often utilise ``basic'' neutron star properties such as mass,
radius and rotation rate to parameterize the mode frequency
\cite{1998MNRAS.299.1059A, 2008PhRvD..78f4063G, 2013PhRvD..88d4052D} but more
recently also more complex quantities such as the effective compactness or tidal
deformability have been employed in the universal relations
\cite{2005ApJ...629..979L, 2010ApJ...714.1234L, 2014PhRvD..90l4023C,
2015PhRvD..92l4004D} in order to improve the accuracy of the estimate. In our
study, we will mostly focus on the most recently proposed universal relation
in Ref.~\cite{Kruger:2019zuz}, as it is the first fitting formula for $f$-mode
frequencies that does not rely on simplifications in their determination.

In this work we adopt a Bayesian framework that allows us to quantify the connection between future measurements of the co- and counterrotating $l=|m|=2$ $f$-modes with key neutron star properties, such as their stellar structure and underlying EOS. This particular choice for the modes is motivated from earlier studies that predict them to be more relevant for typical observations than other modes of the spectrum. The Bayesian framework is related to the one presented recently in Ref.~\cite{Volkel:2020daa}, where it was applied to reconstructing parametrized black hole space-times from their quasi-normal mode spectrum.

Gravitational wave observations of different types of neutron star systems may also come along with complementary information on some of the system properties: e.g., the remnant mass after a neutron star merger may be estimated from the inspiral signal; or the neutron star spin known from radio observation may be accompanied by glitch induced GW signal \cite{2020PhRvD.101j3009H}. Hence, we incorporate mass and spin into our framework. More specifically, we study scenarios in which different levels of prior knowledge on the neutron star mass or independent measurements of the rotation rate are available.

Because the extraction of $f$-modes from complex numerical simulations are computationally time consuming and difficult to automatize, we calculate the $f$-mode frequencies for a sufficiently dense set of equilibrium configurations across the possible parameter space of each considered EOS (they are the same as those used in Ref.~\cite{Kruger:2019zuz}) and interpolate the frequencies of intermediate neutron star models. We call this framework, which always requires an initial choice for the underlying EOS, in the following the {\it EOS method}. The second, complementary approach to the inverse problem is based on using universal relations (UR), which have been reported in the same work \cite{Kruger:2019zuz}. URs allow to compute $f$-mode frequencies from analytic functions of key neutron star parameters up to percent level. The URs have been constructed by fitting the extracted $f$-mode frequencies to a simple analytical function. We call this approach in the following the {\it UR method}.

We find major advantages and disadvantages between the two methods, which are
strongly related to the specific context and what additional information on
neutron star parameters is known. One difference is that the EOS method has
two parameters to uniquely compute the $f$-modes for a given EOS, while the UR
method requires three parameters to do so. If informed prior knowledge on the
remnant mass or additional constraints on the rotation rate are known, meaning
one already knows them to some extent, the UR method provides reliable
constraints on neutron star bulk properties when both $f$-modes have been
observed. Since the currently available universal relation we are using is not
directly incorporating the neutron star radius, it can not directly be used to
recover all neutron star parameters, for which some knowledge of the EOS is
still needed, but most of them. The EOS method has the advantage of being, in
principle, independent of approximations and directly provides all neutron
star parameters that have been computed for the equilibrium solutions
previously. The disadvantage is that the true EOS is not known and assuming a
wrong EOS, i.e., one that does not reflect the physical reality, will in
general yield biased parameters; in our case, our method will yield a neutron
star with different radius and rotation rate as the underlying one which we
used to generate the $f$-mode frequencies (see Sec.~\ref{app_EOS}). The case,
in which two neutron stars constructed using different EOS and stellar structure yield similar $f$-mode frequencies, simply shows that the unique reconstruction requires in general additional information. 

This work is structured as follows. Sec.~\ref{Theory_Methods} provides an overview of the theory and methods being used, which are then applied in Sec.~\ref{applications}. We discuss our findings in Sec.\ref{Discussion} and conclude in Sec.~\ref{Conclusions}.

\section{Theory and Methods}\label{Theory_Methods}

In the following we summarize our methods being used in this work. We start with a review of the underlying theoretical framework to study neutron star oscillations in Sec.~\ref{mathematical_formulation}, before we outline the actual computation of $f$-modes in the present work in Sec.~\ref{interp_f}. We then describe the adopted universal relations in Sec.~\ref{subsec_ur} and the Bayesian analysis in Sec.~\ref{Bayesian_framework}.

\subsection{Mathematical formulation}\label{mathematical_formulation}

The framework in which we calculate the $f$-mode frequencies is laid out in detail in a previous article (see Ref. \cite{PhysRevD.102.064026}), however, for completeness, we will repeat the fundamentals here. In this work we assume standard general relativity (with units in which $G=c=1$), whose field equations, along with the law for the conservation of energy-momentum, are given in terms of the Einstein tensor $G_{\mu \nu}$ and the energy-momentum tensor $T_{\mu \nu}$ as

\begin{align}
	G_{\mu \nu} = 8 \pi T_{\mu \nu}
	\qquad\text{and}\qquad
	\nabla_\mu T^{\mu \nu} = 0.
\end{align}
As is common practice in mode studies, we restrict ourselves to linear perturbations around equilibrium.

To model the neutron star, we consider a perfect fluid with energy density $\epsilon$ and pressure $p$; the energy momentum tensor then takes the form
\begin{align}
	T^{\mu \nu} = \left( \epsilon + p\right) u^{\mu} u^{\nu} + p g^{\mu \nu},
\end{align}
where $u^\mu$ is the 4-velocity and $g^{\mu \nu}$ the underlying metric. Its line element, which describes an axisymmetric neutron star, can be written in quasi-isotropic coordinates as 
\begin{align}
	\text{d}s^2
		 = &-e^{2\nu} \text{d}t^2 + e^{2\psi} r^2 \sin^2 \theta \left(\text{d}\varphi  - \omega \text{d} t\right)^2 \nonumber \\
		& + e^{2\mu} \left(\text{d}r^2 + r^2 \text{d}\theta^2 \right).
\end{align}

In order to construct neutron star equilibrium solutions (for which we employ
the \texttt{rns}-code \cite{1995ApJ...444..306S, 1998A&AS..132..431N, rns-v1.1}), it is necessary to close the system of equations by providing an equation of state that relates $p$ and $\epsilon$. Since the true EOS is still part of current research and subject to large uncertainties, we resort to commonly used proposals for realistic EOSs for our simulations, which are based on various approaches such as detailed microscopic calculations, relativistic mean-field theory, or Skyrme models. In particular, we will use the proposed piecewise-polytropic approximations \cite{2009PhRvD..79l4032R} of the EOSs APR4, H4, MPA1, MS1, SLy4, and WFF1. While different observational constraints, such as a lower bound for the maximum mass \cite{2010Natur.467.1081D} or the radius of a neutron star \cite{2020NatAs...4..625C}, may rule out certain EOSs, our choice of EOSs is intended to cover a broad range of the parameter space.

The field equations then are expanded up to linear order around an equilibrium solution and then evolved in time. While the full oscillation problem has been studied for many decades, and many different family of modes and relations are known from theoretical computations, we here only focus on some part of the so-called $f$-mode spectrum.

\subsection{Computation of $f$-mode frequencies}\label{interp_f}

While our linear perturbation code allows the determination of $f$-mode frequencies at comparatively low computational expense (when compared to non-linear simulations in full general relativity), that task is still tedious and requires manual tweaking of parameters. Hence, we determine the mode frequencies for a sufficiently large number of equilibrium configurations with rotation rates up to the Kepler limit and use linear interpolation to estimate the frequencies of an arbitrary model (based on the same EOS) in the following way: For each of the considered EOS, we work with a tabulated grid which contains the $f$-mode frequencies as function of the star's gravitational mass $M$ and its equatorial radius $R$. Note that in the rotating case, for a given one-parameter EOS, the star's rotation rate $\Omega$ and other parameters are uniquely related to $M$ and $R$. Thus, one is left with a two-dimensional interpolation to relate a specific star with its $f$-mode frequencies. For the technical reasons described above, our grids have a resolution that can in practice not easily be increased arbitrarily. However, we performed several MCMC analysis, beyond the precision reported later in this work, and find that our grids would need to be refined if $f$-modes with below percent precision are studied.

\subsection{Universal relations}\label{subsec_ur}

The estimation of $f$-mode frequencies, based on interpolation as described in the previous Sec.~\ref{interp_f}, are EOS dependent and rely on the availability of sufficiently dense tabulated input for the interpolation. While this can in principle be done for specific choices for EOS, we also study EOS independent properties, commonly known as universal relations. These are obtained by scaling key stellar properties with oscillation modes or damping times and allow to constrain, some, but not all stellar parameters. 

In the following, we provide such an alternative approach by utilizing numerically fitted universal relations in order to compute the frequencies of the co- and counterrotating $f$-modes. Such a relation has been proposed in Ref.~\cite{Kruger:2019zuz} and is given by 

\begin{align}\label{universal_relation}
\hat{\sigma}^{i} =  \left(c^{i}_1 + c^{i}_2 \hat{\Omega} + c^{i}_3\hat{\Omega}^2 \right) +  \left(d^{i}_1 + d^{i}_3 \hat{\Omega}^2 \right) \eta.
\end{align}

Here $i$ denotes the branch of the $f$-mode (co- or counterrotating) and $\hat{\sigma}^{i} = \bar{M} \sigma^{i}/\mathrm{kHz}$ and $\hat{\Omega} = \bar{M} \Omega/\mathrm{kHz}$, where $\bar{M} = M/M_{\astrosun}$. Furthermore, $\eta$ is the effective compactness, which is related to the mass and moment of inertia $I$ of the star via $\eta = \sqrt{\bar{M}^3/I_{45}}$, with $I_{45} = I/10^{45} \mathrm{g \, cm^2}$. The numerical values of the coefficients have been reported in the same work and were obtained by fitting a range of different realistic EOS with stellar sequences \cite{Kruger:2019zuz}. 

The simplicity of universal relations does not only enhance the qualitative understanding of parameter dependencies, but it also allows for back-of-the-envelope estimates. The trade-off is that universal relations are not exact, but come with an intrinsic uncertainty due to their simplicity; in this case the error is at percent level as long as $\Omega$ is not too close to the Kepler limit (see Ref.~\cite{Kruger:2019zuz}). Also note that the universal relation comes at the price of requiring three neutron star parameters, whereas the EOS dependent method is satisfied with only two of them; in a certain way, the EOS is encoded as the third parameter in the universal relation.

\subsection{Bayesian framework}\label{Bayesian_framework}

In the following we briefly summarize the basics of the Bayesian analysis as used in this work. It is similar to the analysis for black hole QNMs, which one of the authors reported recently in Ref.~\cite{Volkel:2020daa}.

Bayes' theorem  connects the parameters $\theta$ of a model with the observed data $D$ via

\begin{align}
P\left(\theta | D \right) = \frac{ P \left(D | \theta \right) P\left(\theta \right)}{P\left( D \right)}.
\end{align}

Here $P\left(\theta | D \right)$ is the posterior, which describes the probability distribution of the parameters given the data. 
It is equal to the likelihood $P \left(D | \theta \right)$, the probability distribution of the data given the parameters, times the prior $P\left(\theta \right)$, the probability distribution of the parameters before looking at the data. The normalization is given by the evidence $P\left( D \right)$, which is the probability of the data itself. In our work we will utilize different levels of informed/uninformed priors for the parameters, which for us here means the distributions are described by narrow/wide Gaussians.

The $f$-mode doublet is the quantity which we consider as data. 
Without real analyzed data, one has to make some assumptions for the likelihood. 
In the following we assume that a future experiment would provide the modes and that the likelihood can be described by two Normal distributions $\mathcal{N}(\mu_i, \sigma_i)$ with  $\mu_i = \nu$ and $\sigma_i = \nu \delta_\nu$. Here $\delta_\nu$ is the dimensionless relative error with which an experiment has determined the $f$-mode frequency $\nu$ (in order to avoid confusion with the standard deviation, we have denoted the $f$-mode frequency here with $\nu$). It can be seen as a free parameter that can be used to qualitatively study the prospects of future gravitational wave detectors.

The actual computation of the posteriors is done using a Markov chain Monte Carlo (MCMC) analysis based on the Python probabilistic programming framework \textsc{PyMC3} \cite{pymc3}. 
The MCMC requires to compute $f$-mode frequencies in each step of the chain; since this is, as mentioned above, not a trivial task, we interpolate it from our data set and couple it via a custom theano (Python library) function to the workflow of \textsc{PyMC3}.

\section{Applications}\label{applications}

In this section, we apply the EOS and UR methods to hypothetically observed $f$-mode frequencies coming from a sample of representative neutron stars with the EOS described previously. In all cases we assume that the co- and counterrotating $f$-mode frequencies are Normal distributed, whereas the one sigma relative error and the prior knowledge on $M$ are being varied depending on the application. Since detailed computations on how precise $f$-mode frequencies can be reconstructed from future observations are highly non-trivial, we assume 3\%-5\%, which seems to be in reach, at least for future detectors, e.g., the Einstein Telescope (see Ref.~\cite{Pratten:2019sed} for a recent study). First, we start with the EOS method in Sec.~\ref{app_EOS} and then apply the UR method in Sec.~\ref{app_UR}. We provide results for their joint application in Sec.~\ref{app_EOS_UR}. 

\subsection{EOS Method}\label{app_EOS}

The EOS method requires the assumption of a particular EOS which we deem to be the physically ``correct'' one to infer the stellar parameters with the MCMC. We have studied this for multiple EOS and varied the priors of the parameters. In the following, we discuss two representative cases. 

\begin{figure}
	\centering
	\includegraphics[width=1.0\linewidth]{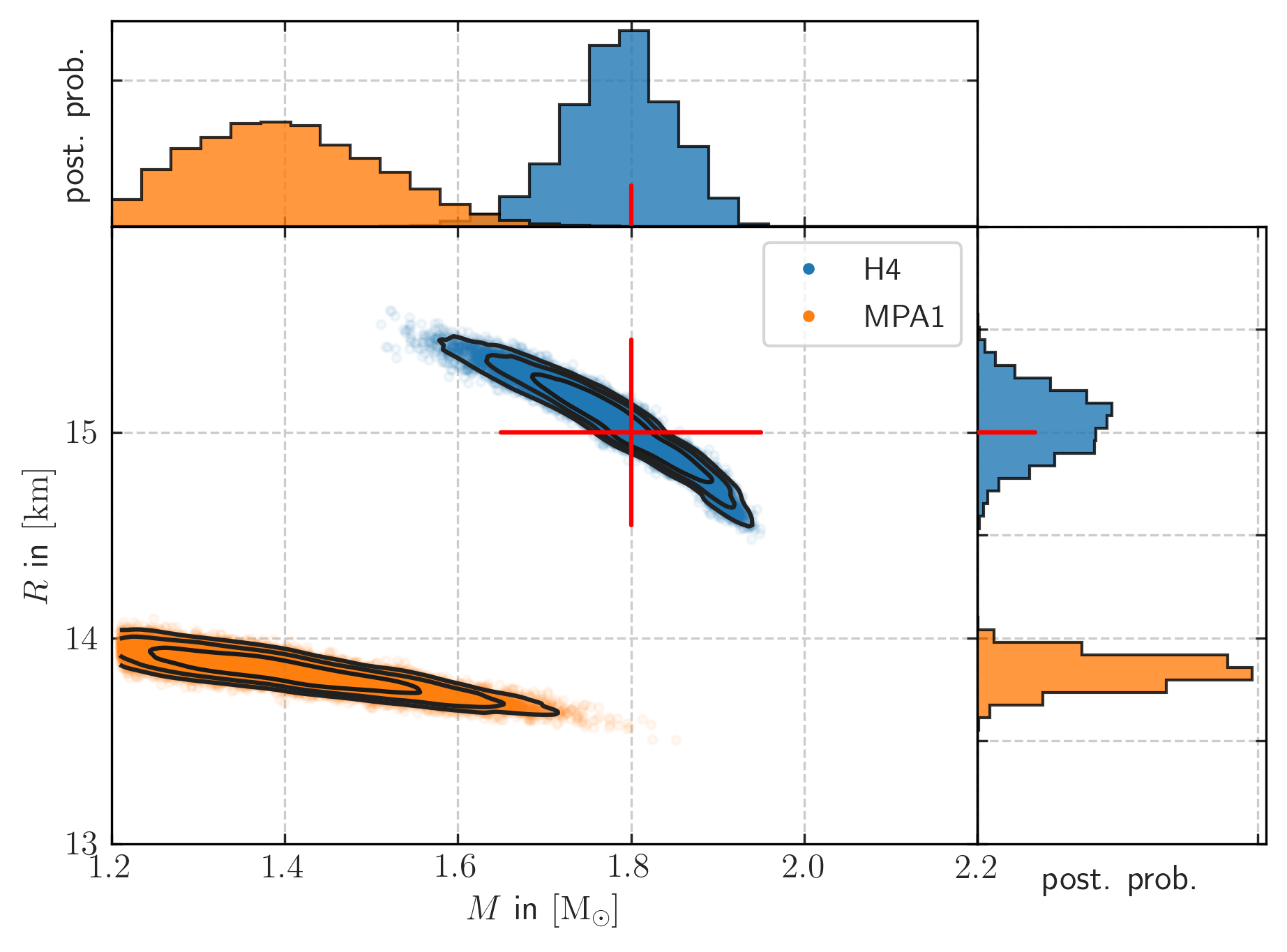}
	\caption{We compare the EOS method assuming the H4 EOS (blue) and the MPA1 EOS (orange). The diagonal panels show the sampled posterior distribution of $M$ and $R$, while the main panel combines a scatter plot with logarithmic contour lines. The red cross and red line indicate the true H4 parameters that belong to the assumed $f$-mode data being used for both EOS for the parameter estimation. The $f$-mode relative error is assumed to be $3\%$.}
	\label{FIG_eos_methods}
\end{figure}

In the first application, which can also be seen as proof of principle, we
chose the H4 EOS and take the $f$-mode frequencies from our data set for a
typical neutron star with $M = 1.8 M_\odot$ and $R = 15\,$km;\footnote{We could also have used two typical values for the $f$-mode frequencies rather than using our data set to simulate a proper neutron star model, however, in that case, we would not be able to test our method by comparing how well it reconstructs the chosen neutron star.} this fully specifies the neutron star and hence all its bulk parameters. These are the true neutron star parameters that we want to infer from the ``observed'' $f$-modes by {\it using the same EOS.} The result, which is represented by blue data, is shown in Fig.~\ref{FIG_eos_methods} and demonstrates that the framework gives reasonable results. The red cross and lines indicate our chosen values for $M$ and $R$.

In the second application, we use the same $f$-modes of the first application, but now assume a different EOS to infer the neutron star parameters. This reflects a less optimistic situation, since there is no hint in the observation that would point towards only one unique EOS. In a ``lucky'' case, it might be such that many of the hitherto proposed EOSs do not support an equilibrium configuration that possesses the specifically observed $f$-modes and would thus be ruled out. However, in general, we do not expect such constraints from observed $f$-mode frequencies as the possible range is rather large; instead, we would expect to recover biased parameters describing a different neutron star with coinciding $f$-mode frequencies. 

We find and report a case in which an alternative EOS yields biased results, which is represented by the orange data in
Fig.~\ref{FIG_eos_methods}. Here we have used the MPA1 EOS for the parameter
estimation. Looking only at the posteriors, there is no indication as to which
of the two EOS is the ``correct'' one (i.e. the one we have selected to
generate the frequencies), or if maybe yet another EOS would explain the
observation better. In both applications we have assumed that $M$ and $R$ have uninformed priors, and that both $f$-modes have a relative error of $3\%$.

Furthermore we verified that the expected distributions of $f$-modes from both EOS are in agreement with the initially provided ones. This has been done by computing their distributions from draws of both of the here shown posteriors. We find no significant deviation that could be used to easily rule out the ``false'' EOS. 

Finally, we also studied how prior knowledge of $M$ changes the above findings. Since precise knowledge of $M$ reduces the given inverse problem to finding the unknown equatorial radius $R$ from two mode frequencies, one would expect that two different EOS become distinguishable. For the above example we report that knowing $M$ within $10\%$ causes strong tensions between posterior and prior of $M$ for the MPA1 EOS, as well as discrepancies for the simulated versus observed $f$-modes. We have verified this by sampling $f$-modes using the posterior distributions of both EOS and compared these with the ones describing the observed $f$-modes. Thus, as expected, one can distinguish the correct from the false EOS. 

\subsection{Universal Relation Method}\label{app_UR}

In the following, we demonstrate the application of the UR method to a range of different pairs of $f$-mode frequencies, representing neutron star models employing different EOSs. Since the UR method is (intentionally) blind to the underlying EOS, but only depends on a subset of the neutron star parameters, the application is straightforward.

\begin{figure}
	\centering
	\includegraphics[width=1.0\linewidth]{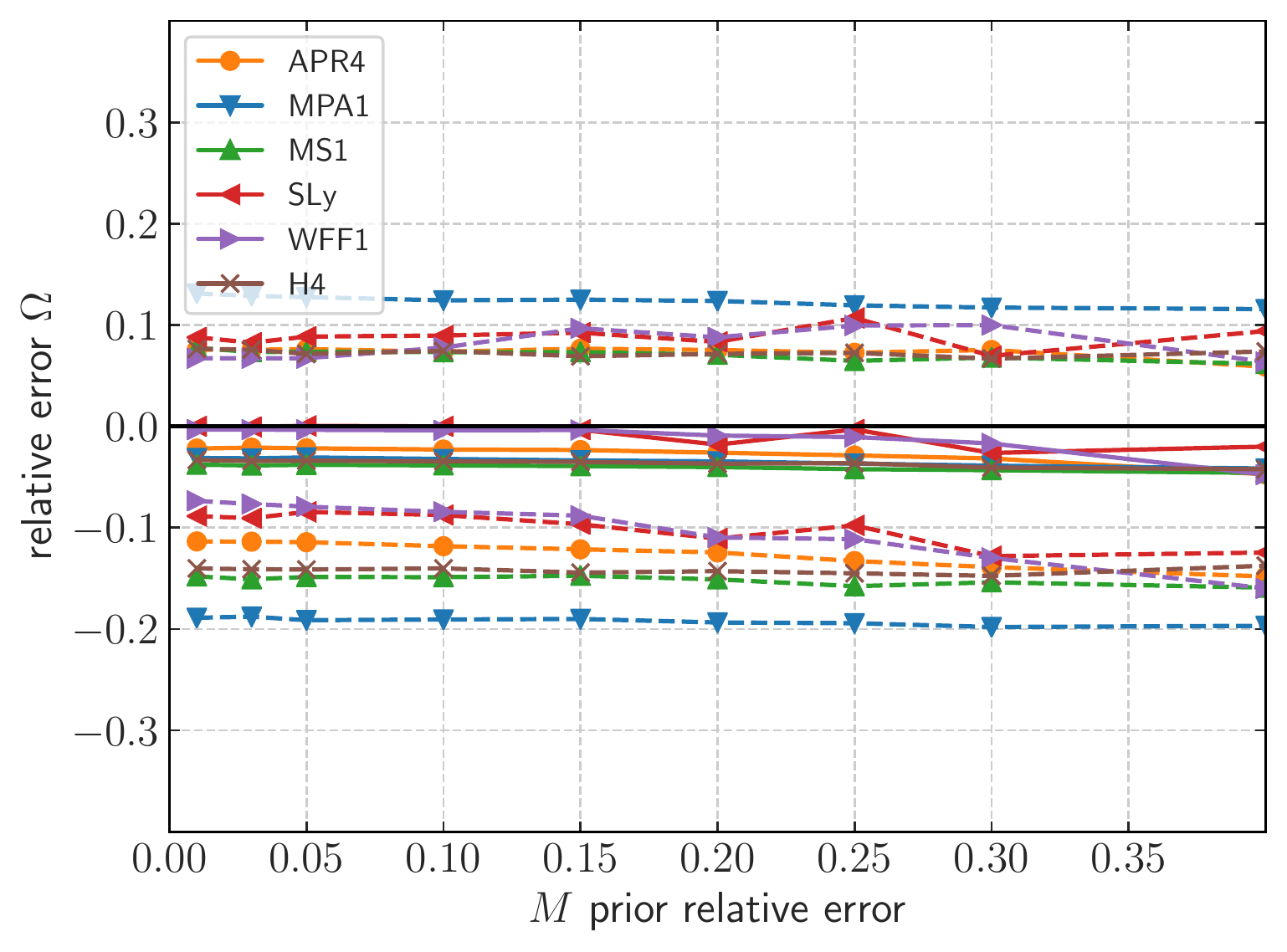}
	\includegraphics[width=1.0\linewidth]{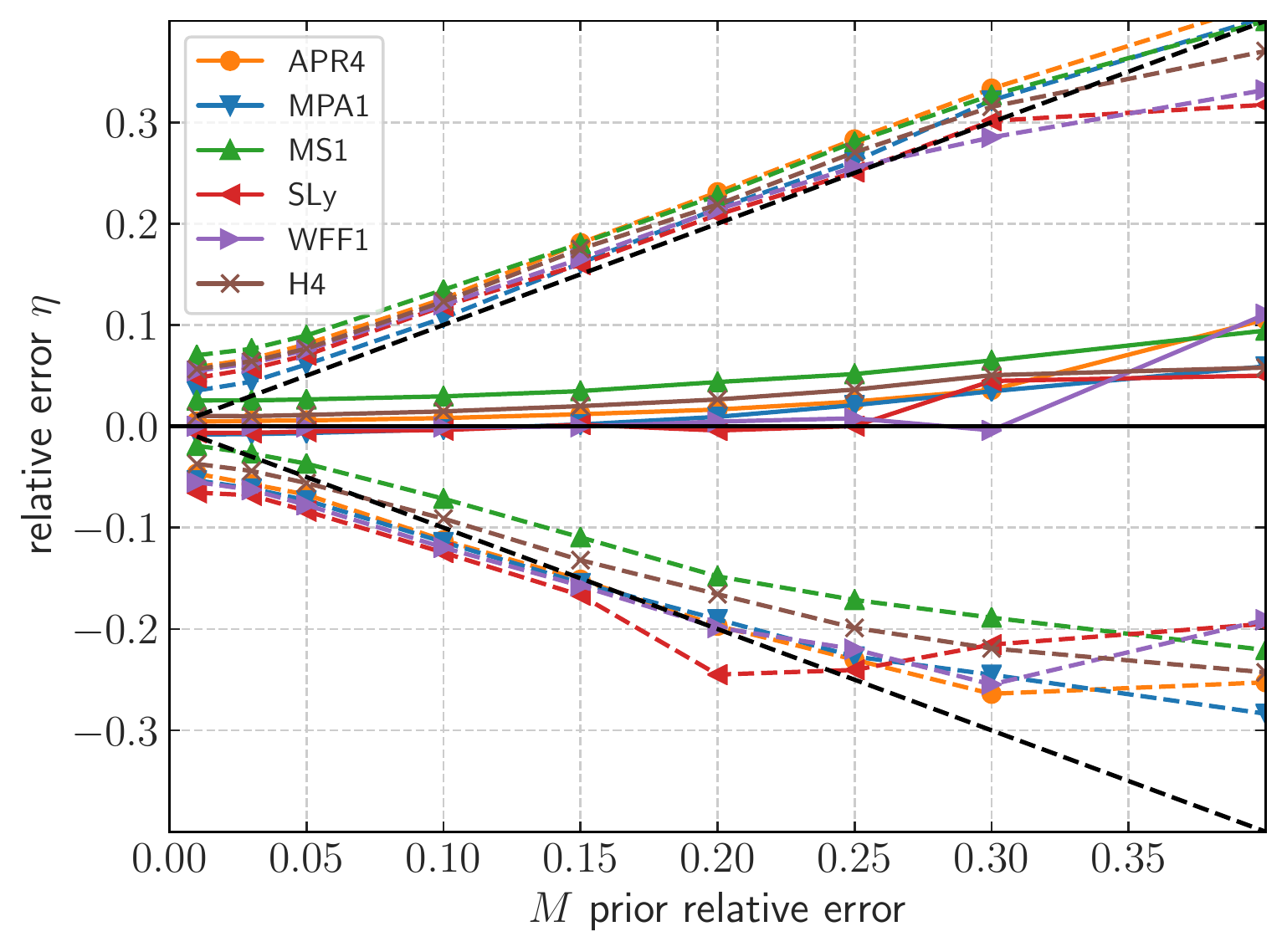}
	\caption{Here we show the relative error of the reconstructed rotation
    rate $\Omega$ (top panel) and effective compactness $\eta$ (bottom panel)
    as function of the relative error of the mass $M$ for different EOS
    (different colors). The dashed lines are the linear interpolation of the
    individual points. The central (solid) lines are the mean value of the reconstruction, while the upper and lower lines show the $68\%$ highest credible interval. The black dashed lines indicate a uniform linear scaling. The $f$-mode relative error is assumed to be $5\%$.}
	\label{FIG_UR}
\end{figure}

We start with the reconstruction of the rotation rate $\Omega$ and its
relative error. Both are shown as a function of the prior
knowledge of $M$ in Fig.~\ref{FIG_UR}. First, it seems that the relative errors of $\Omega$ are almost independent of the prior knowledge of the mass $M$. Second, the absolute values of the relative errors depend more strongly on the underlying EOS than on the knowledge of $M$. However, in all cases it seems to be possible to constrain the rotation rate $\Omega$, for the given $5\%$ precision for the $f$-modes, almost independently from the prior knowledge of $M$ to within $10\%$ to $20\%$.

In the bottom panel of Fig.~\ref{FIG_UR}, we show the corresponding analysis
for the effective compactness $\eta$, again as function of the standard deviation of the prior Normal
distribution for the mass $M$.
Since we assume only two $f$-mode frequencies as observation, but the UR is a function of three parameters, one can only expect to constrain some part of the parameter space. However, if the mass is known to within a few or tens of percent, the parameter space is already strongly confined. We find bounds on $\eta$ which scale for a wide range roughly linear with the uncertainty of $M$. This scaling is comparable for all of the considered EOSs, which suggests that this scaling is also universal.

In order to compare how the two very different scalings depend on the precision of the provided $f$-modes, we repeated the above analysis with smaller relative errors of $3\%$. As expected, the qualitative scaling remains unchanged, but now provides a bit more stringent bounds.

\subsection{Joint EOS and UR Method}\label{app_EOS_UR}

While the EOS method can in principle recover all neutron star parameters, we already argued that the result will, in general, be biased and resemble a different neutron star with similar $f$-modes. In order to quantify how robustly the EOS method is able to constrain the same parameters that can be inferred from the UR method, we have injected a particular $f$-mode pair and applied both methods. The observed $f$-modes have been produced using the H4 EOS, and the EOS method reconstruction is then done via the ``correct'' H4 EOS, the ``wrong'' MPA1 EOS, as well as the UR method. 

\begin{figure}
	\centering
	\includegraphics[width=1.0\linewidth]{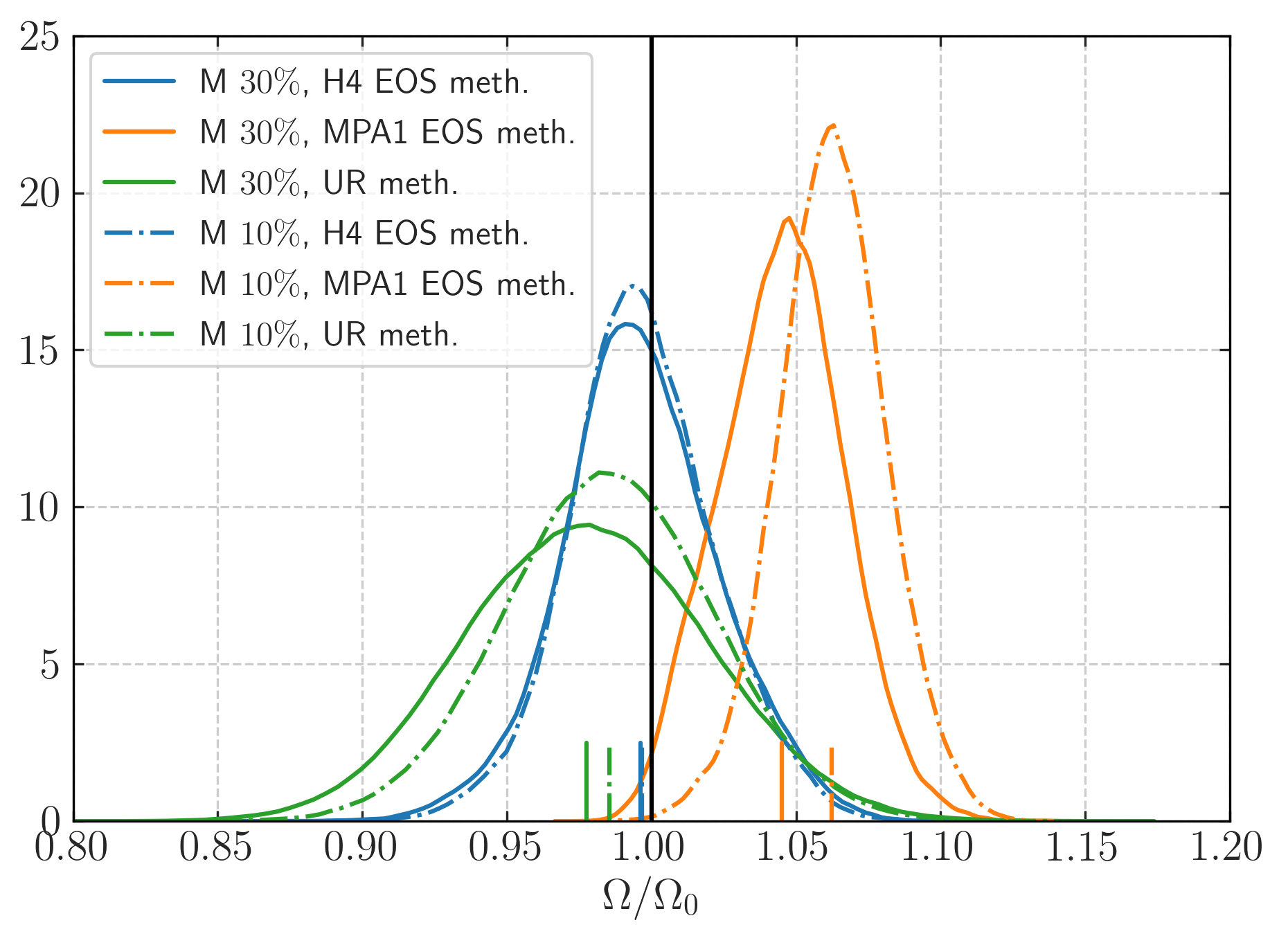}
	\includegraphics[width=1.0\linewidth]{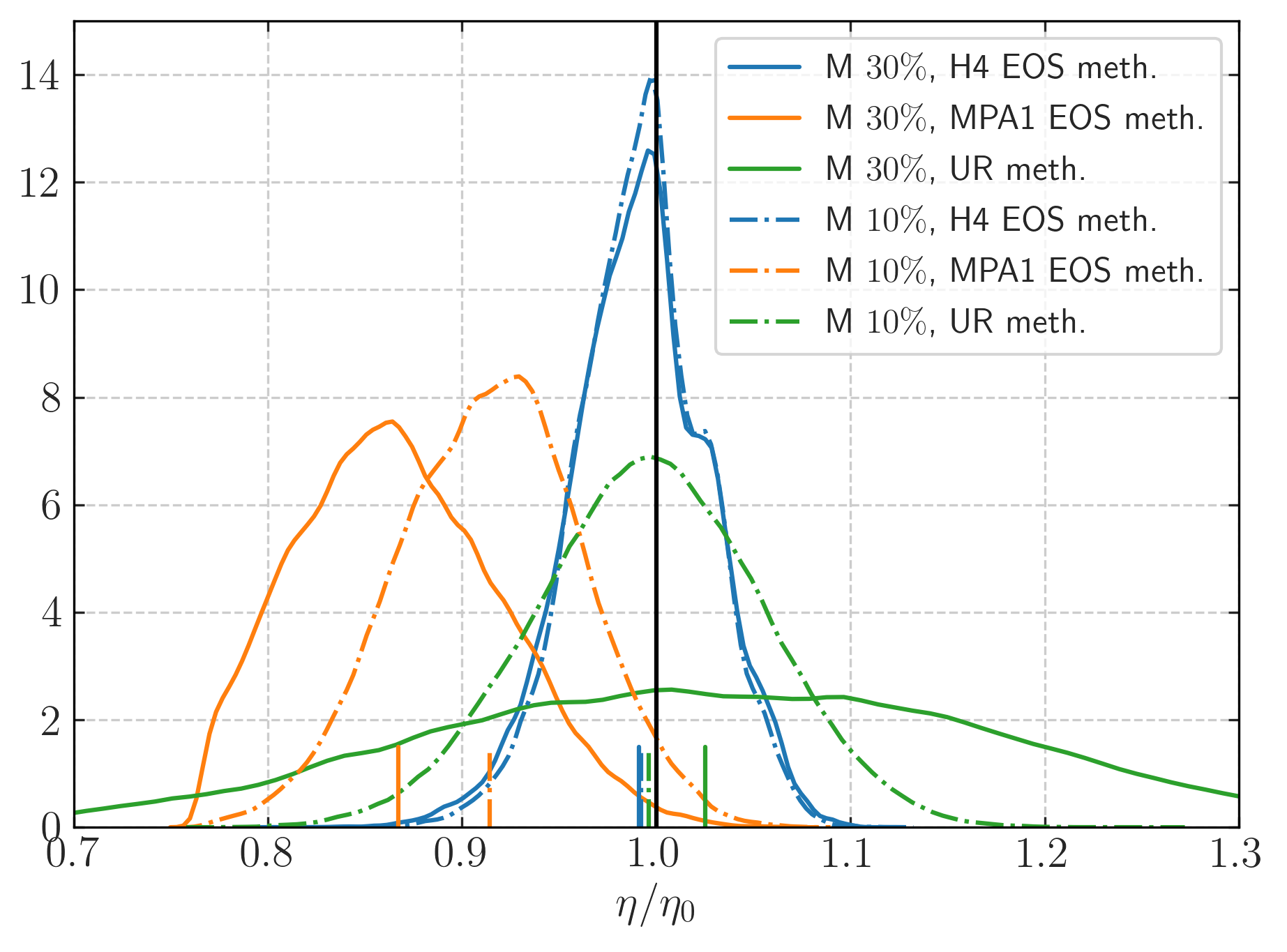}
	\caption{Here we show the posterior distributions of the rotation rate $\Omega$ (top panel) and effective compactness $\eta$ (bottom panel) normalized to the injected H4 values ($\Omega_0, \eta_0$). Posteriors are obtained by using the EOS method with H4 EOS (blue), the MPA1 EOS (orange), as well as the UR method (green). Solid lines correspond to $30\%$ relative error on the prior mass $M$ and dashed lines to $10\%$. We indicate each mean of the posteriors as vertical lines. The $f$-mode relative error is assumed to be $3\%$.}
\label{FIG_eos_vs_ur}
\end{figure}

We report our findings in Fig.~\ref{FIG_eos_vs_ur}, which shows the posterior distribution of the normalized rotation rate $\Omega$ and effective compactness $\eta$ for different prior knowledge on $M$. The normalization is with respect to the true H4 EOS values for the provided $f$-modes. 

It is evident that the H4 EOS method (blue lines) and UR method (green lines) yield very similar results for the rotation rate $\Omega$,
while assuming the MPA1 EOS (orange lines) indicates a value that is larger than the correct one. Note that both observations hold independent
of the specific prior knowledge of $M$ assumed here ($30\%$ and $10\%$).

The situation for the effective compactness $\eta$ is qualitatively different. First, the prior knowledge of $M$ plays a big role for the UR method, but is less important for the EOS methods. For those we find that the correctly assumed H4 EOS is almost independent of uncertainties in $M$, while the posterior distribution obtained by the MPA1 EOS is shifted. Note that the rather different scaling behavior of the UR method is in agreement with the findings of Sec.~\ref{app_UR}.

Finally, while the posteriors of $\Omega$ are very smooth, one observes small
``bumps'' for the H4 EOS, e.g. at $\eta/\eta_0 \approx 1.02$. We have verified that this does not originate from a too small sample size of the MCMC sampling, but most likely is an artifact from the finite resolution and particular range of the used H4 $f$-mode data, as described in Sec.~\ref{interp_f}. This directly sets the scale of how precise our currently implemented EOS data can be used to resolve the underlying parameters, which is of order percent level.

\section{Discussion}\label{Discussion}

In the following we discuss our EOS and UR method based findings in Sec.~\ref{discussion_comparison_EOS_ur}, comment on the uniqueness of the inverse problem in Sec.~\ref{discussion_uniquness}, provide some computational details in Sec.~\ref{discussion_computational}, and provide a brief outlook in Sec.~\ref{discussion_outlook}.

\subsection{EOS and UR methods}\label{discussion_comparison_EOS_ur}

Knowing the correct underlying EOS for hypothetically observed $f$-modes is a very optimistic assumption, since the current variety of proposed realistic EOS still produces quite different neutron stars and $f$-mode spectra. Consequently, by assuming we know the exact EOS one would naively expect to find overly optimistic and biased bounds on the reconstructed stellar parameters. By applying the EOS and UR methods in different contexts in Sec.~\ref{applications}, we have quantified several related aspects and discuss them in the following.

In order to quantify the expectation of biased neutron star parameters, we
applied the EOS method twice to the same pair of $f$-mode frequencies in
Sec.~\ref{app_EOS}. In the first case the same EOS was assumed, in the second
case a different realistic EOS was chosen. The correlations and posteriors for the neutron star mass $M$ and the equatorial radius $R$, both presented in Fig.~\ref{FIG_eos_methods}, clearly show bias. Both reconstructed parameters differ by order $10\%$, and most importantly, the posterior distributions do not overlap significantly. This particular example demonstrates that EOS based reconstruction using interpolation schemes and MCMC is in principle very powerful, but only reliable if the correct EOS is known. Results from any realistic application where the underlying EOS is not known have thus to be interpreted with great care, even if the shape of the posteriors do not show any obvious flaws. 

Since the UR method relies on a simple analytic function with previously fitted coefficients, it can be applied very efficiently. We therefore used it to infer the effective compactness and rotation rate using $f$-modes provided from various realistic EOS and furthermore assumed a wide range of different prior knowledge of the mass in Sec.~\ref{app_UR}. For the effective compactness we find a roughly EOS independent scaling reaching up to few percent level reconstruction of $\eta$, assuming the mass is known with similar precision. However, the posterior distribution of the rotation rate $\Omega$ is almost independent of the prior on $M$ and slightly more EOS dependent. Because the UR method can only be used to infer a combination of mass $M$, effective compactness $\eta$ and rotation rate $\Omega$, it can---by construction---not be used to directly infer the equatorial radius $R$. The most important finding is the robust and thus mostly EOS independent reconstruction of $\eta$ and $\Omega$. This implies that those two parameters can be reliably extracted and their accuracy is mainly limited by the precision with which the $f$-modes can be measured, as well as the prior of $M$.

In the third application, provided in Sec.~\ref{app_EOS_UR}, we directly compared the reconstruction of effective compactness and rotation rate using the same $f$-modes with the two EOS methods from Sec.~\ref{app_EOS} and the UR method. As is evident from Fig.~\ref{FIG_eos_vs_ur} the latter one yields results comparable to those obtained by assuming the correct EOS. For the rotation rate we confirm what can partially already be expected from the results in Sec.~\ref{app_UR}. The UR method includes the correct value, but with larger uncertainties than those coming from the correct EOS method and are quasi independent of the prior of $M$. Since the wrong EOS method seems to converge towards a value being larger than the correct one, but without obvious flaws, one finds biased results, similar to the ones for $M$ and $R$ reported in Sec.~\ref{app_EOS}.

\subsection{Uniqueness of the Inverse Problem}\label{discussion_uniquness}

Unlike in the non-rotating case, the relation between mass and radius is in general not unique. It will crucially depend on the rotation rate that deforms the neutron star, which implies that there are now two unknowns that need to be determined from observations. Measuring two $f$-mode frequencies allows, in principle for any given one-parameter EOS, to uniquely determine the stellar structure. However, this system of equations is not over determined and can thus not be used to rule out certain EOS, unless the observed $f$-modes naturally can not be explained for any neutron star model of that EOS. The consequence is that for the given information, the inverse problem is not uniquely solvable, and in some cases, the biased parameters simply describe viable neutron stars with different structure properties, but very similar $f$-mode spectrum. Only if additional observations, e.g. the mass or rotation rate, can be provided with high accuracy, it is possible to distinguish among the remaining EOS. Especially the robust reconstruction of the rotation rate seems to indicate that glitch induced GW signals could be particularly valuable.

\subsection{Computational aspects}\label{discussion_computational}

Since realistic $f$-mode computations, as well as Bayesian analysis in terms of a MCMC analysis are both computationally expensive we want to make a few comments. 
Reliably and robustly extracting $f$-mode frequencies from a single time
evolution, as done in Ref.~\cite{Kruger:2019zuz}, requires at least a few
dozens of hours on a regular workstation, as well as human interaction to guarantee the robustness of the extracted modes. Having a sufficiently dense parameter space for a specific EOS is therefore cumbersome. However, once it has been obtained, applying the interpolation scheme described in Sec.~\ref{interp_f} allows for almost instant mode computation. Even faster than this is the simple analytic formula describing the universal relation. Depending on the exact size of this parameter space, we find that the UR method speeds up the MCMC analysis by at least one to two orders of magnitude compared to a given EOS method. A typical analysis with sufficient sample size will take a few minutes for the UR method, and at maximum up to a few hours for the EOS method, both on a regular workstation. Especially the application to numerically involved parameter estimation problems demonstrates the enormous advantages of UR based approaches.

\subsection{Outlook}\label{discussion_outlook}

Our analysis only focused on the observation of the co- and counterrotating $f$-mode frequencies, as well as on various choices for informed priors being motivated from different observational scenarios. Other, in principle, available and related quantities are the associated damping times as well as overtones or higher order modes of the acoustic mode spectrum. It can be expected that the reconstruction of stellar parameters would improve if additional modes, e.g. $r$-modes or $g$-modes, were observed, and also the potential to exclude certain EOS. This is particularly interesting for the UR method. As demonstrated in Sec.~\ref{app_UR}, one requires an informed prior on $M$ to obtain reconstructed properties. We argue that this is mainly due to the fact that the universal relations Eq.~\eqref{universal_relation} involve three unknowns $M, \Omega$ and $\eta$. Thus, any additional mode or damping time would lead to at least as many observables as unknowns and therefore, in principle, to a unique reconstruction, modulo intrinsic uncertainties. Another interesting extension of our work could be to perform a Bayesian model comparison to further quantify our observations from Sec.~\ref{app_EOS}, namely how much prior knowledge on neutron star parameters is necessary to distinguish different EOS in cases where they include similar $f$-modes.

\section{Conclusions}\label{Conclusions}

In this work we have studied the inverse problem of rotating neutron stars with realistic equations of state by assuming that measurements of the co- and counterrotating $f$-mode frequencies become available with next generation gravitational wave detectors. These frequencies, along with universal relations, have been reported recently in Ref.~\cite{Kruger:2019zuz,PhysRevD.102.064026}, which defines the theoretical framework of this work. To solve the inverse problem we have conducted a Bayesian analysis by performing MCMC simulations. We provide results using two complementary methods, each of them coming with their own strong and weak aspects. 

The EOS method assumes that the underlying EOS is known and is used to recover the stellar parameters. To perform the computationally expensive analysis, we compute $f$-mode frequencies by interpolating from previously obtained multidimensional tables. The UR method is EOS independent and purely based on universal relations. This allows a fully analytic computation of $f$-modes and thus major computational advantages, which are beneficial for a Bayesian analysis. In order to account for different astrophysical scenarios, in which prior knowledge of the neutron star mass $M$ could be obtained from complementary observations, we study different cases of informed priors for $M$. Furthermore we have selected various representative neutron star models of different EOS, computed their $f$-mode spectrum, and applied the UR method.

The main findings of this work are the following. The UR method is powerful in the reconstruction of the effective compactness, depending on the prior knowledge of $M$, and yields EOS independent results up to a few percent. For all of the here studied cases the reconstructed rotation rate is only mildly depending on the prior of $M$ and its posterior distribution is including the correct value. By construction, the UR method can not be used to directly recover the equatorial radius. The latter one can only be reconstructed using the EOS method by assuming a specific EOS or by performing further calculations. Here we report that if the correct EOS is assumed, the one used to produce the observed $f$-modes, the reconstruction is well behaved and converges towards the injected stellar parameters. However, using a different EOS can in general point towards biased parameters, whose posterior distributions do not necessarily show any flaws, but can admit some shifts once more precise data is included. The falsification of realistic EOS, unless some extreme values of neutron star $f$-modes are considered, will in general require additional and very informed knowledge on other stellar parameters, e.g., the rotation rate or the mass. Finally, we expect that further inclusion of additional universal relations will provide a quantitative and computationally feasible framework to solve the inverse problem of rotating neutron stars.

~\\
\begin{acknowledgments}
SV acknowledges financial support provided under the European Union's H2020 ERC Consolidator Grant ``GRavity from Astrophysical to Microscopic Scales'' grant agreement no. GRAMS-815673. CK acknowledges financial support by DFG research Grant No. 413873357.
\end{acknowledgments}

\bibliography{literature}

\begin{thebibliography}{41}%
\makeatletter
\providecommand \@ifxundefined [1]{%
 \@ifx{#1\undefined}
}%
\providecommand \@ifnum [1]{%
 \ifnum #1\expandafter \@firstoftwo
 \else \expandafter \@secondoftwo
 \fi
}%
\providecommand \@ifx [1]{%
 \ifx #1\expandafter \@firstoftwo
 \else \expandafter \@secondoftwo
 \fi
}%
\providecommand \natexlab [1]{#1}%
\providecommand \enquote  [1]{``#1''}%
\providecommand \bibnamefont  [1]{#1}%
\providecommand \bibfnamefont [1]{#1}%
\providecommand \citenamefont [1]{#1}%
\providecommand \href@noop [0]{\@secondoftwo}%
\providecommand \href [0]{\begingroup \@sanitize@url \@href}%
\providecommand \@href[1]{\@@startlink{#1}\@@href}%
\providecommand \@@href[1]{\endgroup#1\@@endlink}%
\providecommand \@sanitize@url [0]{\catcode `\\12\catcode `\$12\catcode
  `\&12\catcode `\#12\catcode `\^12\catcode `\_12\catcode `\%12\relax}%
\providecommand \@@startlink[1]{}%
\providecommand \@@endlink[0]{}%
\providecommand \url  [0]{\begingroup\@sanitize@url \@url }%
\providecommand \@url [1]{\endgroup\@href {#1}{\urlprefix }}%
\providecommand \urlprefix  [0]{URL }%
\providecommand \Eprint [0]{\href }%
\providecommand \doibase [0]{http://dx.doi.org/}%
\providecommand \selectlanguage [0]{\@gobble}%
\providecommand \bibinfo  [0]{\@secondoftwo}%
\providecommand \bibfield  [0]{\@secondoftwo}%
\providecommand \translation [1]{[#1]}%
\providecommand \BibitemOpen [0]{}%
\providecommand \bibitemStop [0]{}%
\providecommand \bibitemNoStop [0]{.\EOS\space}%
\providecommand \EOS [0]{\spacefactor3000\relax}%
\providecommand \BibitemShut  [1]{\csname bibitem#1\endcsname}%
\let\auto@bib@innerbib\@empty
\bibitem [{\citenamefont {{Abbott}}\ \emph
  {et~al.}(2017{\natexlab{a}})\citenamefont {{Abbott}} \emph
  {et~al.}}]{2017PhRvL.119p1101A}%
  \BibitemOpen
  \bibfield  {author} {\bibinfo {author} {\bibfnamefont {B.~P.}\ \bibnamefont
  {{Abbott}}} \emph {et~al.},\ }\href {\doibase 10.1103/PhysRevLett.119.161101}
  {\bibfield  {journal} {\bibinfo  {journal} {\prl}\ }\textbf {\bibinfo
  {volume} {119}},\ \bibinfo {eid} {161101} (\bibinfo {year}
  {2017}{\natexlab{a}})}\BibitemShut {NoStop}%
\bibitem [{\citenamefont {{Abbott}}\ \emph
  {et~al.}(2017{\natexlab{b}})\citenamefont {{Abbott}} \emph
  {et~al.}}]{2017ApJ...848L..12A}%
  \BibitemOpen
  \bibfield  {author} {\bibinfo {author} {\bibfnamefont {B.~P.}\ \bibnamefont
  {{Abbott}}} \emph {et~al.},\ }\href {\doibase 10.3847/2041-8213/aa91c9}
  {\bibfield  {journal} {\bibinfo  {journal} {\apjl}\ }\textbf {\bibinfo
  {volume} {848}},\ \bibinfo {eid} {L12} (\bibinfo {year}
  {2017}{\natexlab{b}})}\BibitemShut {NoStop}%
\bibitem [{\citenamefont {{Abbott}}\ \emph
  {et~al.}(2017{\natexlab{c}})\citenamefont {{Abbott}} \emph
  {et~al.}}]{2017ApJ...848L..13A}%
  \BibitemOpen
  \bibfield  {author} {\bibinfo {author} {\bibfnamefont {B.~P.}\ \bibnamefont
  {{Abbott}}} \emph {et~al.},\ }\href {\doibase 10.3847/2041-8213/aa920c}
  {\bibfield  {journal} {\bibinfo  {journal} {\apjl}\ }\textbf {\bibinfo
  {volume} {848}},\ \bibinfo {eid} {L13} (\bibinfo {year}
  {2017}{\natexlab{c}})}\BibitemShut {NoStop}%
\bibitem [{\citenamefont {{Abbott}}\ \emph {et~al.}(2020)\citenamefont
  {{Abbott}} \emph {et~al.}}]{2020ApJ...892L...3A}%
  \BibitemOpen
  \bibfield  {author} {\bibinfo {author} {\bibfnamefont {B.~P.}\ \bibnamefont
  {{Abbott}}} \emph {et~al.},\ }\href {\doibase 10.3847/2041-8213/ab75f5}
  {\bibfield  {journal} {\bibinfo  {journal} {\apjl}\ }\textbf {\bibinfo
  {volume} {892}},\ \bibinfo {eid} {L3} (\bibinfo {year} {2020})}\BibitemShut
  {NoStop}%
\bibitem [{\citenamefont {{Margalit}}\ and\ \citenamefont
  {{Metzger}}(2017)}]{2017ApJ...850L..19M}%
  \BibitemOpen
  \bibfield  {author} {\bibinfo {author} {\bibfnamefont {B.}~\bibnamefont
  {{Margalit}}}\ and\ \bibinfo {author} {\bibfnamefont {B.~D.}\ \bibnamefont
  {{Metzger}}},\ }\href {\doibase 10.3847/2041-8213/aa991c} {\bibfield
  {journal} {\bibinfo  {journal} {\apjl}\ }\textbf {\bibinfo {volume} {850}},\
  \bibinfo {eid} {L19} (\bibinfo {year} {2017})}\BibitemShut {NoStop}%
\bibitem [{\citenamefont {{Bauswein}}\ \emph {et~al.}(2017)\citenamefont
  {{Bauswein}}, \citenamefont {{Just}}, \citenamefont {{Janka}},\ and\
  \citenamefont {{Stergioulas}}}]{2017ApJ...850L..34B}%
  \BibitemOpen
  \bibfield  {author} {\bibinfo {author} {\bibfnamefont {A.}~\bibnamefont
  {{Bauswein}}}, \bibinfo {author} {\bibfnamefont {O.}~\bibnamefont {{Just}}},
  \bibinfo {author} {\bibfnamefont {H.-T.}\ \bibnamefont {{Janka}}}, \ and\
  \bibinfo {author} {\bibfnamefont {N.}~\bibnamefont {{Stergioulas}}},\ }\href
  {\doibase 10.3847/2041-8213/aa9994} {\bibfield  {journal} {\bibinfo
  {journal} {\apjl}\ }\textbf {\bibinfo {volume} {850}},\ \bibinfo {eid} {L34}
  (\bibinfo {year} {2017})}\BibitemShut {NoStop}%
\bibitem [{\citenamefont {{Baker}}\ \emph {et~al.}(2017)\citenamefont
  {{Baker}}, \citenamefont {{Bellini}}, \citenamefont {{Ferreira}},
  \citenamefont {{Lagos}}, \citenamefont {{Noller}},\ and\ \citenamefont
  {{Sawicki}}}]{2017PhRvL.119y1301B}%
  \BibitemOpen
  \bibfield  {author} {\bibinfo {author} {\bibfnamefont {T.}~\bibnamefont
  {{Baker}}}, \bibinfo {author} {\bibfnamefont {E.}~\bibnamefont {{Bellini}}},
  \bibinfo {author} {\bibfnamefont {P.~G.}\ \bibnamefont {{Ferreira}}},
  \bibinfo {author} {\bibfnamefont {M.}~\bibnamefont {{Lagos}}}, \bibinfo
  {author} {\bibfnamefont {J.}~\bibnamefont {{Noller}}}, \ and\ \bibinfo
  {author} {\bibfnamefont {I.}~\bibnamefont {{Sawicki}}},\ }\href {\doibase
  10.1103/PhysRevLett.119.251301} {\bibfield  {journal} {\bibinfo  {journal}
  {\prl}\ }\textbf {\bibinfo {volume} {119}},\ \bibinfo {eid} {251301}
  (\bibinfo {year} {2017})}\BibitemShut {NoStop}%
\bibitem [{\citenamefont {{Sakstein}}\ and\ \citenamefont
  {{Jain}}(2017)}]{2017PhRvL.119y1303S}%
  \BibitemOpen
  \bibfield  {author} {\bibinfo {author} {\bibfnamefont {J.}~\bibnamefont
  {{Sakstein}}}\ and\ \bibinfo {author} {\bibfnamefont {B.}~\bibnamefont
  {{Jain}}},\ }\href {\doibase 10.1103/PhysRevLett.119.251303} {\bibfield
  {journal} {\bibinfo  {journal} {\prl}\ }\textbf {\bibinfo {volume} {119}},\
  \bibinfo {eid} {251303} (\bibinfo {year} {2017})}\BibitemShut {NoStop}%
\bibitem [{\citenamefont {{Ezquiaga}}\ and\ \citenamefont
  {{Zumalac{\'a}rregui}}(2017)}]{2017PhRvL.119y1304E}%
  \BibitemOpen
  \bibfield  {author} {\bibinfo {author} {\bibfnamefont {J.~M.}\ \bibnamefont
  {{Ezquiaga}}}\ and\ \bibinfo {author} {\bibfnamefont {M.}~\bibnamefont
  {{Zumalac{\'a}rregui}}},\ }\href {\doibase 10.1103/PhysRevLett.119.251304}
  {\bibfield  {journal} {\bibinfo  {journal} {\prl}\ }\textbf {\bibinfo
  {volume} {119}},\ \bibinfo {eid} {251304} (\bibinfo {year}
  {2017})}\BibitemShut {NoStop}%
\bibitem [{\citenamefont {{Radice}}\ \emph {et~al.}(2018)\citenamefont
  {{Radice}}, \citenamefont {{Perego}}, \citenamefont {{Zappa}},\ and\
  \citenamefont {{Bernuzzi}}}]{2018ApJ...852L..29R}%
  \BibitemOpen
  \bibfield  {author} {\bibinfo {author} {\bibfnamefont {D.}~\bibnamefont
  {{Radice}}}, \bibinfo {author} {\bibfnamefont {A.}~\bibnamefont {{Perego}}},
  \bibinfo {author} {\bibfnamefont {F.}~\bibnamefont {{Zappa}}}, \ and\
  \bibinfo {author} {\bibfnamefont {S.}~\bibnamefont {{Bernuzzi}}},\ }\href
  {\doibase 10.3847/2041-8213/aaa402} {\bibfield  {journal} {\bibinfo
  {journal} {\apjl}\ }\textbf {\bibinfo {volume} {852}},\ \bibinfo {eid} {L29}
  (\bibinfo {year} {2018})}\BibitemShut {NoStop}%
\bibitem [{\citenamefont {{Abbott}}\ \emph {et~al.}(2018)\citenamefont
  {{Abbott}} \emph {et~al.}}]{2018PhRvL.121p1101A}%
  \BibitemOpen
  \bibfield  {author} {\bibinfo {author} {\bibfnamefont {B.~P.}\ \bibnamefont
  {{Abbott}}} \emph {et~al.},\ }\href {\doibase 10.1103/PhysRevLett.121.161101}
  {\bibfield  {journal} {\bibinfo  {journal} {\prl}\ }\textbf {\bibinfo
  {volume} {121}},\ \bibinfo {eid} {161101} (\bibinfo {year}
  {2018})}\BibitemShut {NoStop}%
\bibitem [{\citenamefont {{Bauswein}}\ \emph {et~al.}(2014)\citenamefont
  {{Bauswein}}, \citenamefont {{Stergioulas}},\ and\ \citenamefont
  {{Janka}}}]{2014PhRvD..90b3002B}%
  \BibitemOpen
  \bibfield  {author} {\bibinfo {author} {\bibfnamefont {A.}~\bibnamefont
  {{Bauswein}}}, \bibinfo {author} {\bibfnamefont {N.}~\bibnamefont
  {{Stergioulas}}}, \ and\ \bibinfo {author} {\bibfnamefont {H.~T.}\
  \bibnamefont {{Janka}}},\ }\href {\doibase 10.1103/PhysRevD.90.023002}
  {\bibfield  {journal} {\bibinfo  {journal} {\prd}\ }\textbf {\bibinfo
  {volume} {90}},\ \bibinfo {eid} {023002} (\bibinfo {year}
  {2014})}\BibitemShut {NoStop}%
\bibitem [{\citenamefont {{Takami}}\ \emph {et~al.}(2014)\citenamefont
  {{Takami}}, \citenamefont {{Rezzolla}},\ and\ \citenamefont
  {{Baiotti}}}]{2014PhRvL.113i1104T}%
  \BibitemOpen
  \bibfield  {author} {\bibinfo {author} {\bibfnamefont {K.}~\bibnamefont
  {{Takami}}}, \bibinfo {author} {\bibfnamefont {L.}~\bibnamefont
  {{Rezzolla}}}, \ and\ \bibinfo {author} {\bibfnamefont {L.}~\bibnamefont
  {{Baiotti}}},\ }\href {\doibase 10.1103/PhysRevLett.113.091104} {\bibfield
  {journal} {\bibinfo  {journal} {\prl}\ }\textbf {\bibinfo {volume} {113}},\
  \bibinfo {eid} {091104} (\bibinfo {year} {2014})}\BibitemShut {NoStop}%
\bibitem [{\citenamefont {{Andersson}}\ and\ \citenamefont
  {{Kokkotas}}(1996)}]{1996PhRvL..77.4134A}%
  \BibitemOpen
  \bibfield  {author} {\bibinfo {author} {\bibfnamefont {N.}~\bibnamefont
  {{Andersson}}}\ and\ \bibinfo {author} {\bibfnamefont {K.~D.}\ \bibnamefont
  {{Kokkotas}}},\ }\href {\doibase 10.1103/PhysRevLett.77.4134} {\bibfield
  {journal} {\bibinfo  {journal} {\prl}\ }\textbf {\bibinfo {volume} {77}},\
  \bibinfo {pages} {4134} (\bibinfo {year} {1996})}\BibitemShut {NoStop}%
\bibitem [{\citenamefont {{Andersson}}\ and\ \citenamefont
  {{Kokkotas}}(1998)}]{1998MNRAS.299.1059A}%
  \BibitemOpen
  \bibfield  {author} {\bibinfo {author} {\bibfnamefont {N.}~\bibnamefont
  {{Andersson}}}\ and\ \bibinfo {author} {\bibfnamefont {K.~D.}\ \bibnamefont
  {{Kokkotas}}},\ }\href {\doibase 10.1046/j.1365-8711.1998.01840.x} {\bibfield
   {journal} {\bibinfo  {journal} {\mnras}\ }\textbf {\bibinfo {volume}
  {299}},\ \bibinfo {pages} {1059} (\bibinfo {year} {1998})}\BibitemShut
  {NoStop}%
\bibitem [{\citenamefont {{Lattimer}}\ and\ \citenamefont
  {{Schutz}}(2005)}]{2005ApJ...629..979L}%
  \BibitemOpen
  \bibfield  {author} {\bibinfo {author} {\bibfnamefont {J.~M.}\ \bibnamefont
  {{Lattimer}}}\ and\ \bibinfo {author} {\bibfnamefont {B.~F.}\ \bibnamefont
  {{Schutz}}},\ }\href {\doibase 10.1086/431543} {\bibfield  {journal}
  {\bibinfo  {journal} {\apj}\ }\textbf {\bibinfo {volume} {629}},\ \bibinfo
  {pages} {979} (\bibinfo {year} {2005})}\BibitemShut {NoStop}%
\bibitem [{\citenamefont {{Christensen-Dalsgaard}}\ and\ \citenamefont
  {{Gough}}(1976)}]{1976Natur.259...89C}%
  \BibitemOpen
  \bibfield  {author} {\bibinfo {author} {\bibfnamefont {J.}~\bibnamefont
  {{Christensen-Dalsgaard}}}\ and\ \bibinfo {author} {\bibfnamefont {D.~O.}\
  \bibnamefont {{Gough}}},\ }\href {\doibase 10.1038/259089a0} {\bibfield
  {journal} {\bibinfo  {journal} {\nat}\ }\textbf {\bibinfo {volume} {259}},\
  \bibinfo {pages} {89} (\bibinfo {year} {1976})}\BibitemShut {NoStop}%
\bibitem [{\citenamefont
  {{Christensen-Dalsgaard}}(2002)}]{2002RvMP...74.1073C}%
  \BibitemOpen
  \bibfield  {author} {\bibinfo {author} {\bibfnamefont {J.}~\bibnamefont
  {{Christensen-Dalsgaard}}},\ }\href {\doibase 10.1103/RevModPhys.74.1073}
  {\bibfield  {journal} {\bibinfo  {journal} {Reviews of Modern Physics}\
  }\textbf {\bibinfo {volume} {74}},\ \bibinfo {pages} {1073} (\bibinfo {year}
  {2002})}\BibitemShut {NoStop}%
\bibitem [{\citenamefont {{Cowling}}(1941)}]{1941MNRAS.101..367C}%
  \BibitemOpen
  \bibfield  {author} {\bibinfo {author} {\bibfnamefont {T.~G.}\ \bibnamefont
  {{Cowling}}},\ }\href {\doibase 10.1093/mnras/101.8.367} {\bibfield
  {journal} {\bibinfo  {journal} {\mnras}\ }\textbf {\bibinfo {volume} {101}},\
  \bibinfo {pages} {367} (\bibinfo {year} {1941})}\BibitemShut {NoStop}%
\bibitem [{\citenamefont {{McDermott}}\ \emph {et~al.}(1985)\citenamefont
  {{McDermott}}, \citenamefont {{Hansen}}, \citenamefont {{van Horn}},\ and\
  \citenamefont {{Buland}}}]{1985ApJ...297L..37M}%
  \BibitemOpen
  \bibfield  {author} {\bibinfo {author} {\bibfnamefont {P.~N.}\ \bibnamefont
  {{McDermott}}}, \bibinfo {author} {\bibfnamefont {C.~J.}\ \bibnamefont
  {{Hansen}}}, \bibinfo {author} {\bibfnamefont {H.~M.}\ \bibnamefont {{van
  Horn}}}, \ and\ \bibinfo {author} {\bibfnamefont {R.}~\bibnamefont
  {{Buland}}},\ }\href {\doibase 10.1086/184553} {\bibfield  {journal}
  {\bibinfo  {journal} {\apjl}\ }\textbf {\bibinfo {volume} {297}},\ \bibinfo
  {pages} {L37} (\bibinfo {year} {1985})}\BibitemShut {NoStop}%
\bibitem [{\citenamefont {{Kokkotas}}\ and\ \citenamefont
  {{Schutz}}(1992)}]{1992MNRAS.255..119K}%
  \BibitemOpen
  \bibfield  {author} {\bibinfo {author} {\bibfnamefont {K.~D.}\ \bibnamefont
  {{Kokkotas}}}\ and\ \bibinfo {author} {\bibfnamefont {B.~F.}\ \bibnamefont
  {{Schutz}}},\ }\href {\doibase 10.1093/mnras/255.1.119} {\bibfield  {journal}
  {\bibinfo  {journal} {\mnras}\ }\textbf {\bibinfo {volume} {255}},\ \bibinfo
  {pages} {119} (\bibinfo {year} {1992})}\BibitemShut {NoStop}%
\bibitem [{\citenamefont {{Thorne}}\ and\ \citenamefont
  {{Campolattaro}}(1967)}]{1967ApJ...149..591T}%
  \BibitemOpen
  \bibfield  {author} {\bibinfo {author} {\bibfnamefont {K.~S.}\ \bibnamefont
  {{Thorne}}}\ and\ \bibinfo {author} {\bibfnamefont {A.}~\bibnamefont
  {{Campolattaro}}},\ }\href {\doibase 10.1086/149288} {\bibfield  {journal}
  {\bibinfo  {journal} {\apj}\ }\textbf {\bibinfo {volume} {149}},\ \bibinfo
  {pages} {591} (\bibinfo {year} {1967})}\BibitemShut {NoStop}%
\bibitem [{\citenamefont {{Lindblom}}\ and\ \citenamefont
  {{Detweiler}}(1983)}]{1983ApJS...53...73L}%
  \BibitemOpen
  \bibfield  {author} {\bibinfo {author} {\bibfnamefont {L.}~\bibnamefont
  {{Lindblom}}}\ and\ \bibinfo {author} {\bibfnamefont {S.~L.}\ \bibnamefont
  {{Detweiler}}},\ }\href {\doibase 10.1086/190884} {\bibfield  {journal}
  {\bibinfo  {journal} {\apjs}\ }\textbf {\bibinfo {volume} {53}},\ \bibinfo
  {pages} {73} (\bibinfo {year} {1983})}\BibitemShut {NoStop}%
\bibitem [{\citenamefont {{Jones}}\ \emph {et~al.}(2002)\citenamefont
  {{Jones}}, \citenamefont {{Andersson}},\ and\ \citenamefont
  {{Stergioulas}}}]{2002MNRAS.334..933J}%
  \BibitemOpen
  \bibfield  {author} {\bibinfo {author} {\bibfnamefont {D.~I.}\ \bibnamefont
  {{Jones}}}, \bibinfo {author} {\bibfnamefont {N.}~\bibnamefont
  {{Andersson}}}, \ and\ \bibinfo {author} {\bibfnamefont {N.}~\bibnamefont
  {{Stergioulas}}},\ }\href {\doibase 10.1046/j.1365-8711.2002.05566.x}
  {\bibfield  {journal} {\bibinfo  {journal} {\mnras}\ }\textbf {\bibinfo
  {volume} {334}},\ \bibinfo {pages} {933} (\bibinfo {year}
  {2002})}\BibitemShut {NoStop}%
\bibitem [{\citenamefont {{Gaertig}}\ and\ \citenamefont
  {{Kokkotas}}(2008)}]{2008PhRvD..78f4063G}%
  \BibitemOpen
  \bibfield  {author} {\bibinfo {author} {\bibfnamefont {E.}~\bibnamefont
  {{Gaertig}}}\ and\ \bibinfo {author} {\bibfnamefont {K.~D.}\ \bibnamefont
  {{Kokkotas}}},\ }\href {\doibase 10.1103/PhysRevD.78.064063} {\bibfield
  {journal} {\bibinfo  {journal} {\prd}\ }\textbf {\bibinfo {volume} {78}},\
  \bibinfo {eid} {064063} (\bibinfo {year} {2008})}\BibitemShut {NoStop}%
\bibitem [{\citenamefont {{Kr{\"u}ger}}\ and\ \citenamefont
  {{Kokkotas}}(2020{\natexlab{a}})}]{PhysRevD.102.064026}%
  \BibitemOpen
  \bibfield  {author} {\bibinfo {author} {\bibfnamefont {C.~J.}\ \bibnamefont
  {{Kr{\"u}ger}}}\ and\ \bibinfo {author} {\bibfnamefont {K.~D.}\ \bibnamefont
  {{Kokkotas}}},\ }\href {\doibase 10.1103/PhysRevD.102.064026} {\bibfield
  {journal} {\bibinfo  {journal} {\prd}\ }\textbf {\bibinfo {volume} {102}},\
  \bibinfo {eid} {064026} (\bibinfo {year} {2020}{\natexlab{a}})}\BibitemShut
  {NoStop}%
\bibitem [{\citenamefont {{Doneva}}\ \emph {et~al.}(2013)\citenamefont
  {{Doneva}}, \citenamefont {{Gaertig}}, \citenamefont {{Kokkotas}},\ and\
  \citenamefont {{Kr{\"u}ger}}}]{2013PhRvD..88d4052D}%
  \BibitemOpen
  \bibfield  {author} {\bibinfo {author} {\bibfnamefont {D.~D.}\ \bibnamefont
  {{Doneva}}}, \bibinfo {author} {\bibfnamefont {E.}~\bibnamefont {{Gaertig}}},
  \bibinfo {author} {\bibfnamefont {K.~D.}\ \bibnamefont {{Kokkotas}}}, \ and\
  \bibinfo {author} {\bibfnamefont {C.}~\bibnamefont {{Kr{\"u}ger}}},\ }\href
  {\doibase 10.1103/PhysRevD.88.044052} {\bibfield  {journal} {\bibinfo
  {journal} {\prd}\ }\textbf {\bibinfo {volume} {88}},\ \bibinfo {eid} {044052}
  (\bibinfo {year} {2013})}\BibitemShut {NoStop}%
\bibitem [{\citenamefont {{Lau}}\ \emph {et~al.}(2010)\citenamefont {{Lau}},
  \citenamefont {{Leung}},\ and\ \citenamefont {{Lin}}}]{2010ApJ...714.1234L}%
  \BibitemOpen
  \bibfield  {author} {\bibinfo {author} {\bibfnamefont {H.~K.}\ \bibnamefont
  {{Lau}}}, \bibinfo {author} {\bibfnamefont {P.~T.}\ \bibnamefont {{Leung}}},
  \ and\ \bibinfo {author} {\bibfnamefont {L.~M.}\ \bibnamefont {{Lin}}},\
  }\href {\doibase 10.1088/0004-637X/714/2/1234} {\bibfield  {journal}
  {\bibinfo  {journal} {\apj}\ }\textbf {\bibinfo {volume} {714}},\ \bibinfo
  {pages} {1234} (\bibinfo {year} {2010})}\BibitemShut {NoStop}%
\bibitem [{\citenamefont {{Chan}}\ \emph {et~al.}(2014)\citenamefont {{Chan}},
  \citenamefont {{Sham}}, \citenamefont {{Leung}},\ and\ \citenamefont
  {{Lin}}}]{2014PhRvD..90l4023C}%
  \BibitemOpen
  \bibfield  {author} {\bibinfo {author} {\bibfnamefont {T.~K.}\ \bibnamefont
  {{Chan}}}, \bibinfo {author} {\bibfnamefont {Y.~H.}\ \bibnamefont {{Sham}}},
  \bibinfo {author} {\bibfnamefont {P.~T.}\ \bibnamefont {{Leung}}}, \ and\
  \bibinfo {author} {\bibfnamefont {L.~M.}\ \bibnamefont {{Lin}}},\ }\href
  {\doibase 10.1103/PhysRevD.90.124023} {\bibfield  {journal} {\bibinfo
  {journal} {\prd}\ }\textbf {\bibinfo {volume} {90}},\ \bibinfo {eid} {124023}
  (\bibinfo {year} {2014})}\BibitemShut {NoStop}%
\bibitem [{\citenamefont {{Doneva}}\ and\ \citenamefont
  {{Kokkotas}}(2015)}]{2015PhRvD..92l4004D}%
  \BibitemOpen
  \bibfield  {author} {\bibinfo {author} {\bibfnamefont {D.~D.}\ \bibnamefont
  {{Doneva}}}\ and\ \bibinfo {author} {\bibfnamefont {K.~D.}\ \bibnamefont
  {{Kokkotas}}},\ }\href {\doibase 10.1103/PhysRevD.92.124004} {\bibfield
  {journal} {\bibinfo  {journal} {\prd}\ }\textbf {\bibinfo {volume} {92}},\
  \bibinfo {eid} {124004} (\bibinfo {year} {2015})}\BibitemShut {NoStop}%
\bibitem [{\citenamefont {{Kr{\"u}ger}}\ and\ \citenamefont
  {{Kokkotas}}(2020{\natexlab{b}})}]{Kruger:2019zuz}%
  \BibitemOpen
  \bibfield  {author} {\bibinfo {author} {\bibfnamefont {C.~J.}\ \bibnamefont
  {{Kr{\"u}ger}}}\ and\ \bibinfo {author} {\bibfnamefont {K.~D.}\ \bibnamefont
  {{Kokkotas}}},\ }\href {\doibase 10.1103/PhysRevLett.125.111106} {\bibfield
  {journal} {\bibinfo  {journal} {\prl}\ }\textbf {\bibinfo {volume} {125}},\
  \bibinfo {eid} {111106} (\bibinfo {year} {2020}{\natexlab{b}})}\BibitemShut
  {NoStop}%
\bibitem [{\citenamefont {V\"olkel}\ and\ \citenamefont
  {Barausse}(2020)}]{Volkel:2020daa}%
  \BibitemOpen
  \bibfield  {author} {\bibinfo {author} {\bibfnamefont {S.~H.}\ \bibnamefont
  {V\"olkel}}\ and\ \bibinfo {author} {\bibfnamefont {E.}~\bibnamefont
  {Barausse}},\ }\href {\doibase 10.1103/PhysRevD.102.084025} {\bibfield
  {journal} {\bibinfo  {journal} {Phys. Rev. D}\ }\textbf {\bibinfo {volume}
  {102}},\ \bibinfo {pages} {084025} (\bibinfo {year} {2020})}\BibitemShut
  {NoStop}%
\bibitem [{\citenamefont {{Ho}}\ \emph {et~al.}(2020)\citenamefont {{Ho}},
  \citenamefont {{Jones}}, \citenamefont {{Andersson}},\ and\ \citenamefont
  {{Espinoza}}}]{2020PhRvD.101j3009H}%
  \BibitemOpen
  \bibfield  {author} {\bibinfo {author} {\bibfnamefont {W.~C.~G.}\
  \bibnamefont {{Ho}}}, \bibinfo {author} {\bibfnamefont {D.~I.}\ \bibnamefont
  {{Jones}}}, \bibinfo {author} {\bibfnamefont {N.}~\bibnamefont
  {{Andersson}}}, \ and\ \bibinfo {author} {\bibfnamefont {C.~M.}\ \bibnamefont
  {{Espinoza}}},\ }\href {\doibase 10.1103/PhysRevD.101.103009} {\bibfield
  {journal} {\bibinfo  {journal} {\prd}\ }\textbf {\bibinfo {volume} {101}},\
  \bibinfo {eid} {103009} (\bibinfo {year} {2020})}\BibitemShut {NoStop}%
\bibitem [{\citenamefont {{Stergioulas}}\ and\ \citenamefont
  {{Friedman}}(1995)}]{1995ApJ...444..306S}%
  \BibitemOpen
  \bibfield  {author} {\bibinfo {author} {\bibfnamefont {N.}~\bibnamefont
  {{Stergioulas}}}\ and\ \bibinfo {author} {\bibfnamefont {J.~L.}\ \bibnamefont
  {{Friedman}}},\ }\href {\doibase 10.1086/175605} {\bibfield  {journal}
  {\bibinfo  {journal} {\apj}\ }\textbf {\bibinfo {volume} {444}},\ \bibinfo
  {pages} {306} (\bibinfo {year} {1995})}\BibitemShut {NoStop}%
\bibitem [{\citenamefont {{Nozawa}}\ \emph {et~al.}(1998)\citenamefont
  {{Nozawa}}, \citenamefont {{Stergioulas}}, \citenamefont {{Gourgoulhon}},\
  and\ \citenamefont {{Eriguchi}}}]{1998A&AS..132..431N}%
  \BibitemOpen
  \bibfield  {author} {\bibinfo {author} {\bibfnamefont {T.}~\bibnamefont
  {{Nozawa}}}, \bibinfo {author} {\bibfnamefont {N.}~\bibnamefont
  {{Stergioulas}}}, \bibinfo {author} {\bibfnamefont {E.}~\bibnamefont
  {{Gourgoulhon}}}, \ and\ \bibinfo {author} {\bibfnamefont {Y.}~\bibnamefont
  {{Eriguchi}}},\ }\href {\doibase 10.1051/aas:1998304} {\bibfield  {journal}
  {\bibinfo  {journal} {\aaps}\ }\textbf {\bibinfo {volume} {132}},\ \bibinfo
  {pages} {431} (\bibinfo {year} {1998})}\BibitemShut {NoStop}%
\bibitem [{\citenamefont {{Stergioulas}}(1995)}]{rns-v1.1}%
  \BibitemOpen
  \bibfield  {author} {\bibinfo {author} {\bibfnamefont {N.}~\bibnamefont
  {{Stergioulas}}},\ }\href@noop {} {}\bibinfo {howpublished}
  {\url{http://www.gravity.phys.uwm.edu/rns/}} (\bibinfo {year}
  {1995})\BibitemShut {NoStop}%
\bibitem [{\citenamefont {{Read}}\ \emph {et~al.}(2009)\citenamefont {{Read}},
  \citenamefont {{Lackey}}, \citenamefont {{Owen}},\ and\ \citenamefont
  {{Friedman}}}]{2009PhRvD..79l4032R}%
  \BibitemOpen
  \bibfield  {author} {\bibinfo {author} {\bibfnamefont {J.~S.}\ \bibnamefont
  {{Read}}}, \bibinfo {author} {\bibfnamefont {B.~D.}\ \bibnamefont
  {{Lackey}}}, \bibinfo {author} {\bibfnamefont {B.~J.}\ \bibnamefont
  {{Owen}}}, \ and\ \bibinfo {author} {\bibfnamefont {J.~L.}\ \bibnamefont
  {{Friedman}}},\ }\href {\doibase 10.1103/PhysRevD.79.124032} {\bibfield
  {journal} {\bibinfo  {journal} {\prd}\ }\textbf {\bibinfo {volume} {79}},\
  \bibinfo {eid} {124032} (\bibinfo {year} {2009})}\BibitemShut {NoStop}%
\bibitem [{\citenamefont {{Demorest}}\ \emph {et~al.}(2010)\citenamefont
  {{Demorest}}, \citenamefont {{Pennucci}}, \citenamefont {{Ransom}},
  \citenamefont {{Roberts}},\ and\ \citenamefont
  {{Hessels}}}]{2010Natur.467.1081D}%
  \BibitemOpen
  \bibfield  {author} {\bibinfo {author} {\bibfnamefont {P.~B.}\ \bibnamefont
  {{Demorest}}}, \bibinfo {author} {\bibfnamefont {T.}~\bibnamefont
  {{Pennucci}}}, \bibinfo {author} {\bibfnamefont {S.~M.}\ \bibnamefont
  {{Ransom}}}, \bibinfo {author} {\bibfnamefont {M.~S.~E.}\ \bibnamefont
  {{Roberts}}}, \ and\ \bibinfo {author} {\bibfnamefont {J.~W.~T.}\
  \bibnamefont {{Hessels}}},\ }\href {\doibase 10.1038/nature09466} {\bibfield
  {journal} {\bibinfo  {journal} {\nat}\ }\textbf {\bibinfo {volume} {467}},\
  \bibinfo {pages} {1081} (\bibinfo {year} {2010})}\BibitemShut {NoStop}%
\bibitem [{\citenamefont {{Capano}}\ \emph {et~al.}(2020)\citenamefont
  {{Capano}}, \citenamefont {{Tews}}, \citenamefont {{Brown}}, \citenamefont
  {{Margalit}}, \citenamefont {{De}}, \citenamefont {{Kumar}}, \citenamefont
  {{Brown}}, \citenamefont {{Krishnan}},\ and\ \citenamefont
  {{Reddy}}}]{2020NatAs...4..625C}%
  \BibitemOpen
  \bibfield  {author} {\bibinfo {author} {\bibfnamefont {C.~D.}\ \bibnamefont
  {{Capano}}}, \bibinfo {author} {\bibfnamefont {I.}~\bibnamefont {{Tews}}},
  \bibinfo {author} {\bibfnamefont {S.~M.}\ \bibnamefont {{Brown}}}, \bibinfo
  {author} {\bibfnamefont {B.}~\bibnamefont {{Margalit}}}, \bibinfo {author}
  {\bibfnamefont {S.}~\bibnamefont {{De}}}, \bibinfo {author} {\bibfnamefont
  {S.}~\bibnamefont {{Kumar}}}, \bibinfo {author} {\bibfnamefont {D.~A.}\
  \bibnamefont {{Brown}}}, \bibinfo {author} {\bibfnamefont {B.}~\bibnamefont
  {{Krishnan}}}, \ and\ \bibinfo {author} {\bibfnamefont {S.}~\bibnamefont
  {{Reddy}}},\ }\href {\doibase 10.1038/s41550-020-1014-6} {\bibfield
  {journal} {\bibinfo  {journal} {Nature Astronomy}\ }\textbf {\bibinfo
  {volume} {4}},\ \bibinfo {pages} {625} (\bibinfo {year} {2020})}\BibitemShut
  {NoStop}%
\bibitem [{\citenamefont {Salvatier}\ \emph {et~al.}(2016)\citenamefont
  {Salvatier}, \citenamefont {Wiecki},\ and\ \citenamefont
  {Fonnesbeck}}]{pymc3}%
  \BibitemOpen
  \bibfield  {author} {\bibinfo {author} {\bibfnamefont {J.}~\bibnamefont
  {Salvatier}}, \bibinfo {author} {\bibfnamefont {T.~V.}\ \bibnamefont
  {Wiecki}}, \ and\ \bibinfo {author} {\bibfnamefont {C.}~\bibnamefont
  {Fonnesbeck}},\ }\href {\doibase 10.7717/peerj-cs.55} {\bibfield  {journal}
  {\bibinfo  {journal} {PeerJ Computer Science}\ }\textbf {\bibinfo {volume}
  {2}},\ \bibinfo {pages} {e55} (\bibinfo {year} {2016})}\BibitemShut {NoStop}%
\bibitem [{\citenamefont {Pratten}\ \emph {et~al.}(2020)\citenamefont
  {Pratten}, \citenamefont {Schmidt},\ and\ \citenamefont
  {Hinderer}}]{Pratten:2019sed}%
  \BibitemOpen
  \bibfield  {author} {\bibinfo {author} {\bibfnamefont {G.}~\bibnamefont
  {Pratten}}, \bibinfo {author} {\bibfnamefont {P.}~\bibnamefont {Schmidt}}, \
  and\ \bibinfo {author} {\bibfnamefont {T.}~\bibnamefont {Hinderer}},\ }\href
  {\doibase 10.1038/s41467-020-15984-5} {\bibfield  {journal} {\bibinfo
  {journal} {Nature Commun.}\ }\textbf {\bibinfo {volume} {11}},\ \bibinfo
  {pages} {2553} (\bibinfo {year} {2020})},\ \Eprint
  {http://arxiv.org/abs/1905.00817} {arXiv:1905.00817 [gr-qc]} \BibitemShut
  {NoStop}%
\end{thebibliography}%

\end{document}